\documentclass[superscriptaddress, groupedaddress, twocolumn, amsmath,amssymb, aps, prl]{revtex4}
\usepackage{graphicx}
\usepackage{bm}
\usepackage{color}
\usepackage[colorlinks=true ,citecolor=blue,urlcolor=blue]{hyperref}

\def\micron{{\ \mu{\rm m}}}					

\newcommand{\bra}[1]{\langle #1|}
\newcommand{\ket}[1]{|#1\rangle}

\definecolor{coolgrey}{rgb}{0.55, 0.57, 0.67}
\definecolor{bleudefrance}{rgb}{0.19, 0.55, 0.91}
\definecolor{crimson}{rgb}{0.86, 0.08, 0.24}
\definecolor{orangepeel}{rgb}{1.0, 0.62, 0.0}
\definecolor{orange(colorwheel)}{rgb}{1.0, 0.5, 0.0}

\DeclareUnicodeCharacter{202F}{\,}

\begin{document}
	
\title{Realization of a Synthetic Hall Torus with a Topological Charge Pump in a Spinor Bose-Einstein Condensate}

\author{T.-H. Chien}
\author{S.-C. Wu}
\author{Y.-H. Su}
\author{L.-R. Liu}
\author{N.-C. Chiu}
\affiliation{Institute of Atomic and Molecular Sciences, Academia Sinica, Taipei, Taiwan 10617}
\author{M. Sarkar}
\affiliation{Department of Physics, University of Mumbai, Mumbai 400098, India}
\author{Q. Zhou}
\email{zhou753@purdue.edu}
\affiliation{Department of Physics and Astronomy, Purdue University, West Lafayette, Indiana 47907-2036, USA}
\author{Y.-J. Lin}
\email{linyj@as.edu.tw}
\affiliation{Institute of Atomic and Molecular Sciences, Academia Sinica, Taipei, Taiwan 10617}
\affiliation{Department of Physics, National Tsing Hua University, Hsinchu 30013, Taiwan}
	
\date{\today}

\begin{abstract}
	We report the first experimental realization of a synthetic Hall torus using a spinor Bose-Einstein condensate confined in a ring-shaped trap with \textit{in situ} imaging. By cyclically coupling three hyperfine spin states via Raman and microwave fields, we impose a periodic boundary condition in the synthetic dimension, which, together with a real-space ring trap, realizes a toroidal geometry with synthetic magnetic flux penetrating the surface. This flux induces azimuthal density modulations whose periodicity is uniquely determined by the quantized toroidal flux, a hallmark of the Hall torus. We then implement a topological charge pump by adiabatically varying an axial flux through the torus, inducing quantized azimuthal transport of the atoms. We further investigate the onset of these modulations as the system transitions from cylindrical to toroidal topology. Our results establish a versatile platform for exploring quantum Hall physics and topological phenomena in synthetic curved spaces.
\end{abstract}
\maketitle

The development of synthetic dimensions~\cite{Boada2012,Celi2014,Ozawa2019,Fabre2024, buser2020interacting,arguello2024synthetic} using internal atomic states has opened new pathways to simulating higher-dimensional physics and engineering unconventional boundary conditions. Beyond hyperfine spin states~\cite{Stuhl2015,Mancini2015,Chalopin2020,Roell2023,Han2019,Luo2020,Li2022, Anderson2020,Liang2021,Fabre2022,Yan2019,Zhang2021}, synthetic dimensions have been realized using momentum modes~\cite{Meier2016,Meier2016a,Xie2020,An2018,Gou2020}, vibrational states~\cite{salerno2019quantized,price2017synthetic}, and Rydberg atom arrays~\cite{Kanungo2022,Chen2024}, enabling versatile control of synthetic quantum matter. These platforms have revealed phenomena such as chiral edge transport and nontrivial band topology. Notably, periodic boundary conditions in the synthetic dimension enable geometries inaccessible in real space, including synthetic cylinders and tori with magnetic flux penetrating their surfaces. Theoretical work predicts rich band structures and symmetry-protected features in such geometries~\cite{Yan2019}, while experimental synthetic Hall cylinders~\cite{Han2019, Li2022, Anderson2020, Liang2021,Fabre2022} have laid the groundwork for more complex topological structures.

The Hall torus provides a fundamentally important platform for exploring quantum phenomena in curved and topologically nontrivial spaces. In quantum Hall physics, although local properties such as fractional quasiparticles and their braiding phases can be probed in open systems, the defining fingerprints of intrinsic topological order are global and exist only on closed manifolds. For instance, a Laughlin $\nu = 1/q$ state has exactly $q$ degenerate ground states on a torus~\cite{wen1990ground}. 
These degenerate states are indistinguishable by local observables but globally distinct due to long-range quantum entanglement~\cite{kitaev2003fault}.
Threading one flux quantum through a hole of the torus cycles the system within the degenerate manifold~\cite{wen1990ground}. The many-body Chern number defined via twisted boundary conditions in two independent directions is meaningful only on a torus~\cite{niu1985quantized}. These features, unattainable in planar systems, make the Hall torus an ideal setting for probing the interplay between geometry, topology, and quantum coherence. 
Nevertheless, despite progress with Hall-cylinder geometries, an experimental Hall-torus platform has remained unavailable.

In this Letter, we report the first experimental realization of a synthetic Hall torus using an atomic $F=1$ spinor Bose-Einstein condensate confined in a ring-shaped potential. The synthetic dimension is encoded in the three magnetic sublevels $m_F = \pm1, 0$, cyclically coupled by Raman and microwave fields. This cyclic coupling imposes a periodic boundary condition in the synthetic dimension, which, together with the real-space ring trap of radius $R\sim14\micron$, realizes a toroidal geometry [Fig.~1(b)], with the synthetic dimension defining the poloidal direction of the torus. To emulate a Hall torus, we create a synthetic magnetic field penetrating the toroidal surface, corresponding to a transverse flux, using Raman coupling that simultaneously changes the spin state $m_F$ and the center-of-mass orbital angular momentum (OAM) of the atoms.  This scheme imparts a net OAM transfer of $\Delta \ell = 2\hbar$ within the same spin state, producing coherent interference between wavefunctions with different OAM. As a hallmark of the Hall torus, we adopt \textit{in situ} imaging to observe a characteristic density modulation with two distinct minima (or maxima) along the azimuthal direction of the ring, absent when the microwave coupling is turned off that leaves the synthetic dimension with open boundary conditions. We realize a toroidal topological charge pump by adiabatically varying the axial magnetic flux threading the torus in time, yielding quantized azimuthal transport. Finally, by exploiting tunable atom–laser interactions, we dynamically switch the boundary condition in the synthetic dimension and study the nonequilibrium emergence of density modulations as the system acquires toroidal topology.

We start by describing our setup and Hamiltonian. The Raman coupling is implemented with two laser beams copropagating along ${\mathbf e}_z$: a Gaussian beam and a Laguerre-Gaussian (LG) beam with phase winding number 1, carrying OAM $\hbar$. These beams couple bare spin states $\ket{m_F}$ to $\ket{m_F \pm 1}$ while transferring $\pm\hbar$ of OAM [Fig.~1(c)], realizing spin--orbital-angular-momentum coupling~\cite{sun2015spin,chen2016spin,Chen2018,Chen2018a,zhang2019ground}. The Raman coupling strength has an approximately cylindrically symmetric profile $\Omega(r) = \Omega_M \sqrt{e}\,(r/r_M)\,e^{-r^2/2r_M^2}$, whose spatially nonuniform light shift confines atoms near $r_M$, the radial position of maximum coupling, and creates a ring-shaped trapping potential.
In our setup, $\Omega_M/2\pi = 3.8$~kHz and $r_M = 17~\mu\mathrm{m}$. The Raman detuning is $\delta = \Delta\omega_L - \omega_Z$, where $\Delta\omega_L$ is the Raman frequency difference and $\omega_Z$ is the effective linear Zeeman shift. A two-photon microwave field resonantly couples $\ket{m_F = 1}$ and $\ket{m_F = -1}$ with a strength $\Omega_{\rm mw}/2\pi = 3.15\pm0.25$~kHz. At $\delta = 0$, the total atom--light coupling Hamiltonian in cylindrical coordinates $(r,\phi,z)$ is
\begin{align}\label{eq:V}
	V/\hbar &= \frac{\Omega(r)}{\sqrt{2}}\,e^{i\phi} \ket{1}\bra{0} + \frac{\Omega(r)}{\sqrt{2}}\,e^{i\phi} \ket{0}\bra{-1} \nonumber\\
	&\quad + \frac{\Omega_{\rm mw}}{2}\,e^{-i\theta_{\rm mw}} \ket{-1}\bra{1} + \text{H.c.} - \omega_q \ket{0}\bra{0}.
\end{align}
Here $\omega_q$ is the effective quadratic Zeeman shift, and both $\omega_Z$ and $\omega_q$ include microwave-induced light shifts (see Supplemental Material~\cite{supplemental}). We measure $\omega_q/2\pi=1.6$~kHz and 50~Hz for $\Omega_{\rm mw}/2\pi = 3.15$~kHz and 0, respectively. The microwave phase $\theta_{\rm mw}$ is actively locked to the Raman relative phase, ensuring reproducible and tunable density-minimum positions. The Raman couplings $\ket{-1}\rightarrow\ket{0}$ and $\ket{0}\rightarrow\ket{1}$ carry the phase factor $e^{i\phi}$, while the microwave coupling $\ket{1}\rightarrow\ket{-1}$ carries the position-independent phase $-\theta_{\rm mw}$. We express fluxes in phase units, with one flux quantum corresponding to $2\pi$. The microwave phase therefore represents an axial flux $-\theta_{\rm mw}$ threading the torus through the poloidal cycle, analogous to that in a Hall cylinder~\cite{Liang2021,Fabre2022}. The spatially varying Raman phases generate a synthetic magnetic field $B_{\perp}$ penetrating the toroidal surface. For infinitesimal plaquettes bounded by [$\phi,\phi+d\phi$], the accumulated Peierls phase is $d\phi$ on each Raman-coupled strip, $[m_F=-1,0]$ and $[m_F=0,1]$, and zero on the microwave-coupled strip $[m_F=1,-1]$. Thus, each Raman-coupled one-third of the torus surface carries flux $2\pi$, while the remaining one-third carries no transverse flux. The total transverse flux is therefore $4\pi$, corresponding to two flux quanta.

\begin{figure}
	\centering
	\includegraphics[width=3.5 in]{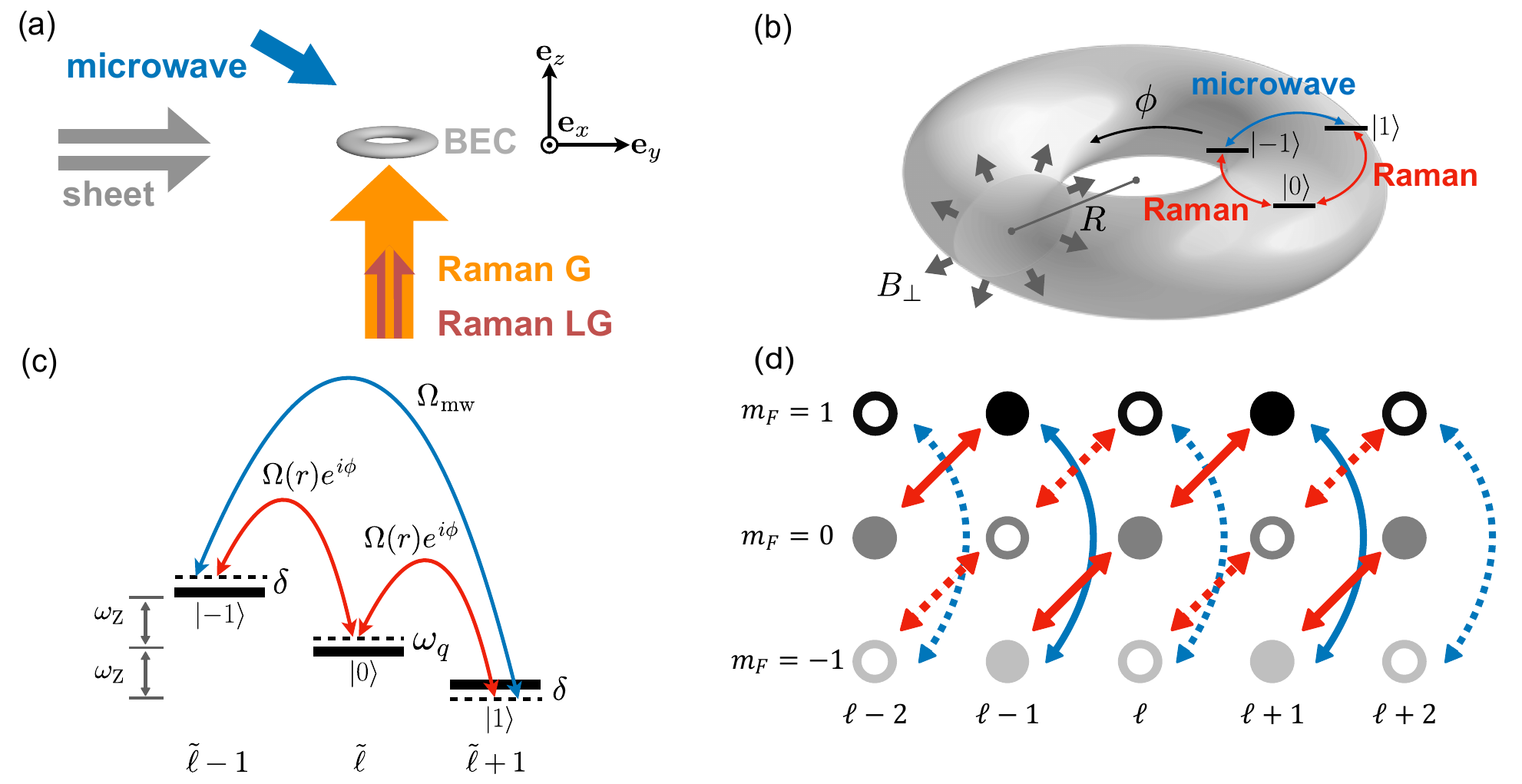}
	\caption{(a) Experimental setup. (b) The torus combines a ring trap in real space with cyclically coupled spin states in the synthetic dimension. (c) Schematic of the cyclic coupling implemented by Raman (red) and microwave fields (blue). (d) Cyclic coupling displayed in the spin and OAM space. Solid (open) symbols denote atoms in torus branch 1(2).}
\end{figure}

The atom-light interaction in Eq.~(\ref{eq:V}) gives the eigenstates a characteristic azimuthal structure.
%
%
Denoting a state with spin $m_F$ and OAM $l$ (in unit of $\hbar$) by $|m_F,l\rangle\propto e^{il\phi}$
, these states are coupled via the sequence $...|-1,l=\ell\rangle \leftrightarrow |0,\ell+1\rangle \leftrightarrow |1,\ell+2\rangle\leftrightarrow|-1,\ell+2\rangle...$ [Fig.~1(d)]. Under cylindrical symmetry, the eigenstates are grouped into two decoupled branches, labeled by quasi-OAM $\ell$ and denoted by filled (branch 1) and hollow (branch 2) circles, in Fig.~1(d). The eigenstates of branch 1, 2 are
\begin{subequations}
	\begin{align}
		|\Phi_1\rangle&=\sum_{m_F=0,\pm1}\sum_{n=-\infty}^{\infty} c_{1,m_F,n}|m_F, \ell+m_F+2n\rangle,\label{eigenstate1}\\
		|\Phi_2\rangle&=\sum_{m_F=0,\pm1}\sum_{n=-\infty}^{\infty} c_{2,m_F,n}|m_F, \ell +m_F+1+2n\rangle,\label{eigenstate2}
	\end{align}
\end{subequations}
where $c_{1,m_F,n}$ and $c_{2,m_F,n}$ gives the weights of the OAM components. Given the kinetic energy $l^2\hbar^2/2mR^2=l^2E_0$, where $E_0\approx h\times 0.3$~Hz is much smaller than $\Omega_{\rm mw}$ and $\Omega_M$, each $|m_F\rangle$ contains multiple OAM states. For each $|m_F\rangle$, these OAM states are equally spaced by $d_\phi=2$, analogous to the reciprocal lattice vector in a linear lattice; this gives an azimuthal lattice period $2\pi/d_\phi=\pi$. Thus the column density, $n_{m_F}^{\rm 2D}(r,\phi)=\int dz n_{m_F}^{\rm 3D}(r,\phi,z)$, satisfies $n^{\rm 2D}_{m_F}(r,\phi)=n^{\rm2D}_{m_F}(r,\phi+\pi)$, and exhibits two maxima and two minima around the ring.
%
%
By contrast, without microwave coupling, each $|m_F\rangle$ is associated with a single OAM state and the density modulation vanishes. Another notable feature is that $\theta_{\rm mw}$ controls the locations of the maxima (minima). Directly seen from Eq.~(\ref{eq:V}), a gauge transformation $|1\rangle\rightarrow e^{i\theta_{\rm mw}/2}|1\rangle $ and $|-1\rangle\rightarrow e^{-i\theta_{\rm mw}/2}|-1\rangle $ eliminates the phase of $|-1\rangle\langle 1|$ while modifying phases of $|1\rangle\langle 0|$ and $|0\rangle\langle -1|$ to $e^{i(\phi-\theta_{\rm mw}/2)}$. Physical quantities such as $n^{\rm 2D}_{m_F}$ therefore depend on $\phi-\theta_{\rm mw}/2$, corresponding to the gauge-invariant cyclic accumulated phase $2\phi-\theta_{\rm mw}$.

\begin{figure}
	\centering
	\includegraphics[width=3.5 in]{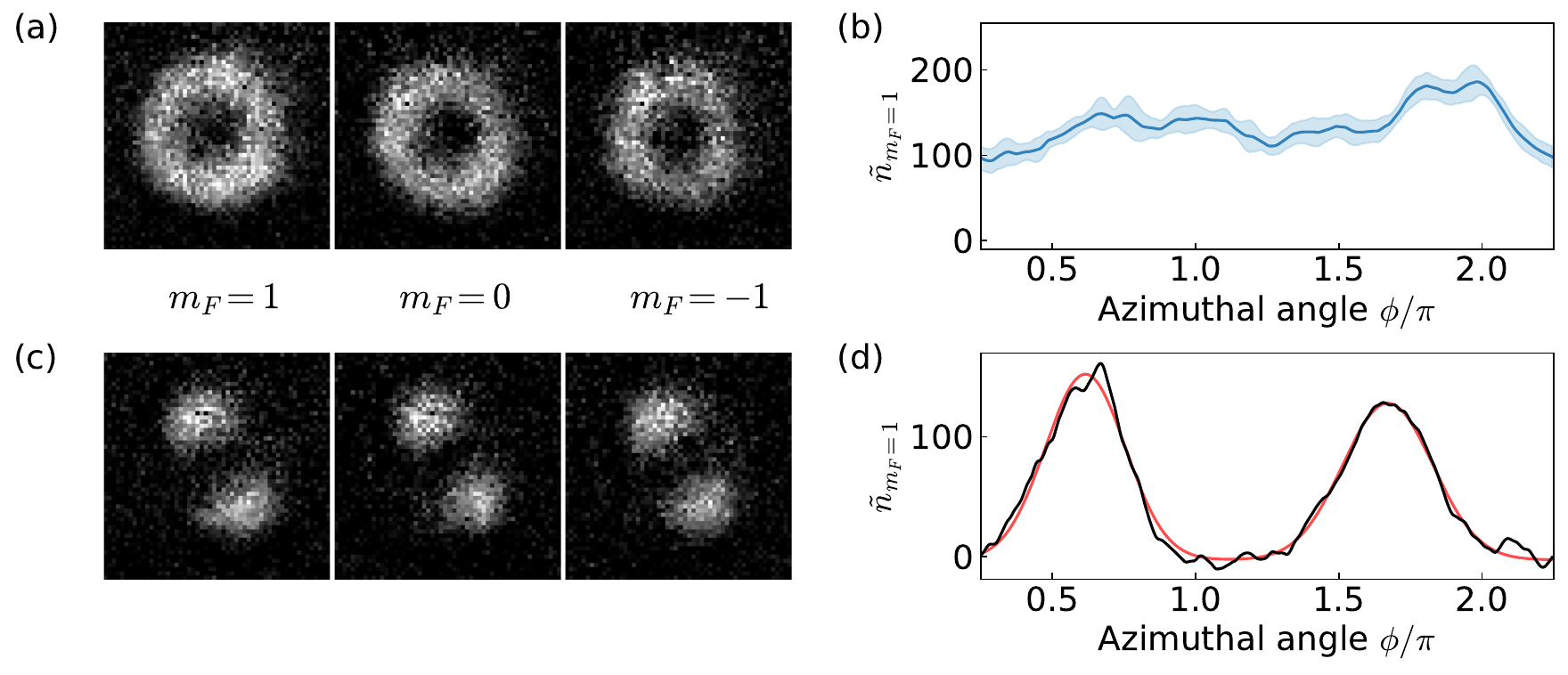}
	\caption{Atomic optical density (OD) of $\ket{m_F=1,0,-1}$ before turning on the microwave coupling $\Omega_{\rm mw}$ (a) and after turning on $\Omega_{\rm mw}$ in 50~ms with $\theta_{\rm mw} = 0$ and $t_h\lesssim 0.1$~ms (c). The color scale for the density profile is given by the maximum value in each panel. The field of view is $72 \times 72 \micron^2$. Azimuthal profiles of the ring (b) and torus (d) are displayed as the effective 1D density $\tilde{n}_{m_F=1}$ vs. $\phi$. (b) shows the average and standard deviation of 10 identical experimental realizations. Black (red) curve in (d) denotes the experimental data (fit). The images in (a) and (c) are single-shot.
	}
\end{figure}

In our experiment, the trapping frequencies along $z$ and $r$ are $f_z\approx370$~Hz and $f_r\approx61$~Hz, respectively, while the chemical potential is $\mu/h\sim$ 600 Hz. The atomic cloud is therefore well described by a mean-field condensate wavefunction occupying a common axial and radial mode. To focus on the azimuthal distribution while accounting for residual fluctuations in $z$ and $r$, we defined an effective 1D density $\tilde{n}_{m_F}(\phi-\theta_{\rm mw}/2)\propto\int dr r n_{m_F}^{\rm 2D}(r,\phi-\theta_{\rm mw}/2)$. The extrema locations of $\tilde{n}_{m_F}$ are thus controlled by $\theta_{\rm mw}$.

Our experiment begins with $^{87}$Rb atoms in $\ket{F = 1, m_F = -1}$ state confined in a crossed optical dipole trap combined with a blue-detuned sheet potential. We adiabatically load cold thermal atoms
into the lowest-energy spinor branch Raman-dressed state $\ket{\xi_{-1}}$~\cite{Chen2018a} by ramping on the Raman coupling over 10 ms at fixed detuning $\delta_i/2\pi = 2.8$ kHz, followed by a 40~ms linear ramp of $\delta$ to zero. After loading, the atoms have temperature $T\gtrsim T_c$ where $T_c$ is the BEC critical temperature. Evaporative cooling is then performed by reducing the dipole trap power to near zero over 0.3~s, yielding a condensate of $N = (8 \pm 1) \times 10^4$ atoms in the Raman-dressed state $\varphi(\vec{r})\ket{\xi_{-1}}=\varphi(\vec{r})\left( e^{i\phi} \frac{1 - \cos\beta}{2},\ -\frac{\sin\beta}{\sqrt{2}}, e^{-i\phi}\frac{1 + \cos\beta}{2} \right)^\mathrm{T}$, where $\varphi(\vec{r})$ is the external wave function and $\tan\beta = \Omega(r)/\delta$~\cite{Chen2018a}. At $\delta=0$, the spinor part of the wave function $\ket{\xi_{-1}}=(0.5,-\sqrt{0.5},0.5)$ in the ring-shaped trapping potential generated by the dressed-state light shift, approximately $-\Omega(r)$ (see Supplemental Material~\cite{supplemental}). Here the Raman-induced ring potential differs from real-space ring traps for bare spin states used previously to study superflow in BECs~\cite{wright2013driving,eckel2014hysteresis}. Additionally, the crossed dipole trap provides transverse confinement with $\omega_r/2\pi=25$~Hz.

In an ideal cylindrically symmetric ring trap, $\varphi(\vec{r})$ contains only a single Fourier component $\tilde{\ell}_j=0$, so $\tilde{n}_{m_F}$ is azimuthally uniform before the microwave field is turned on. In practice, imperfect alignment between the Raman beams and the condensate produces azimuthal roughness and additional Fourier components, $\varphi(\vec{r})\approx \sum_j a_je^{i \tilde{\ell}_j\phi}$ with finite $a_{j\neq 0}$.
In addition, the 0.3~s evaporation time may be insufficient compared to $(E_0/h)^{-1}\sim 3.3$~s where $E_0$ is the smallest single-particle trap energy; so some atoms may not fully reach the $\tilde{\ell}_j=0$ ground state. Both effects produce small azimuthal density variations without nodes. As shown in Fig.~2(ab), we perform \textit{in situ} spin-selective imaging by abruptly switching off the Raman and microwave couplings, together with the dipole trap, after a brief 0.5 ms time-of-flight. The atomic optical density (OD) distribution of each $|m_F\rangle$ component exhibits a ring-shaped profile with radius $\sim 14\micron$, where ${\rm OD}_{m_F}=n^{\rm{2D}}_{m_F} \bar{\sigma}$ and $\bar{\sigma}$ is the imaging cross section. At $\delta=0$, the spin population fractions are approximately 0.25 in $m_F = \pm1$ and 0.5 in $m_F = 0$. Figure~2(b) shows only small variations in $\tilde{n}_{m_F}(\phi)$, indicating the aforementioned cylindrical-symmetry breaking factors do not significantly restructure the density profiles.

To engage the cyclic coupling, we ramp on the two-photon microwave coupling $\Omega_{\rm mw}$ over a time $t_{\rm on}$ with phase $\theta_{\rm mw}$, while simultaneously adjusting the bias magnetic field by 2.1~kHz to compensate for the microwave-induced light shift and maintain $\delta \approx 0$. After a variable hold time $t_h$, we perform \textit{in situ} imaging. Figure~2(cd) shows that for $t_{\rm on} = 50$~ms, $\theta_{\rm mw} = 0$ and $t_h\lesssim 0.1$~ms, the atomic OD exhibits two near-zero minima along the azimuthal direction $\phi$, accompanied by two density maxima. This pronounced modulation signifies the formation of the toroidal geometry. We characterize the torus profile versus $t_{\rm on}$ at short $t_h\lesssim 0.1$~ms and find that it becomes largely unchanged for $t_{\rm on} \gtrsim 35$~ms. We therefore use $t_{\rm on} = 50$~ms as the standard loading duration.

Before and after turning on $\Omega_{\rm mw}$, all three spin components $m_F=\pm1,0$ exhibit the same spatial profiles in both experiment [Fig.~2(a) and 2(c)] and simulation (see Supplemental Material~\cite{supplemental}). This spatial miscibility arises from the small spin-dependent interaction energy, only a few Hz and therefore negligible, compared with the Raman and microwave coupling strengths of $\sim3$~kHz.

For general $t_{\rm on}$, we compute $\tilde{n}_{m_F=1}$ versus $\phi$ and fit it to a double Gaussian function, extracting the two peak locations $\phi = \mu_1, \mu_2$ and their root-mean-square widths, $\sigma_1, \sigma_2$. For the standard $t_{\rm on} = 50$~ms with $\theta_{\rm mw}=0$, 10 identical experimental realizations give $\mu_1 = 0.62 \pm 0.02~\pi$ and $\mu_2 = 1.69 \pm 0.02~\pi$, with $\sigma_1= 0.135 \pm 0.02~\pi,\sigma_2= 0.15 \pm 0.01~\pi$, [Fig.~2(cd)]. The separation $\mu_2 -\mu_1 = 1.07\pi \sim \pi$ is consistent with the expected twofold azimuthal modulation of a torus. We also examine ${\rm OD}_{m_F=1}$ versus $r$ at the peak locations $\phi = \mu_1, \mu_2$, and find the radii of maximal OD $R_1 \sim R_2=R \approx 14\micron$.

\begin{figure}
	\centering
	\includegraphics[width=3.5 in]{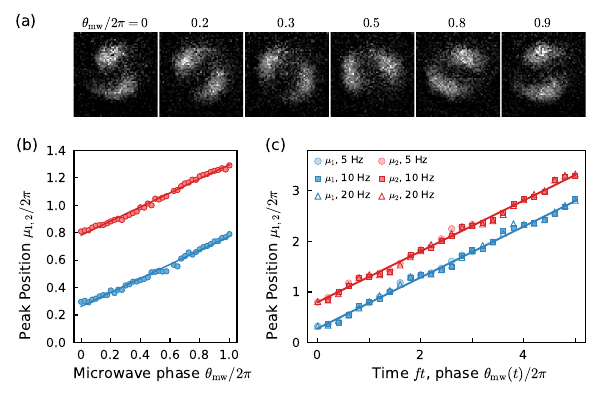}
	\caption{(a) Atomic OD of $\ket{m_F=1}$ vs. microwave phase $\theta_{\rm mw}$ with $t_{\rm on}=50$~ms and $t_h\lesssim 0.1$~ms. The field of view is $72 \times 72 \micron^2$. (b) Blue (red) symbols denote the azimuthal peak position $\mu_1~(\mu_2)$ vs. $\theta_{\rm mw}$. Lines denote the respective linear fits, both with slope $0.51\pm 0.01$. (c) Blue (red) symbols denote $\mu_1~(\mu_2)$ vs. the scaled time $ft=\theta_{\rm mw}(t)/2\pi$. Lines denote the respective linear fits, both with slope $0.50\pm 0.01$.}
\end{figure}

In the ideal cylindrically symmetric case, only one torus branch exists, and the two maxima in $\tilde{n}_{m_F}(\phi)$ are identical. In practice, ring trap imperfectness and finite evaporation time leads to small branch mixing and a minor peak asymmetry.
We quantify this asymmetry using the peak amplitudes $A_{1,2}$ and widths in $\tilde{n}_{m_F=1}$, and the area ratio $r_a = A_1 \sigma_1 / A_2 \sigma_2$, proportional to the atom-number ratio.
For $\theta_{\rm mw}=0$, we find $r_a = 1.1 \pm 0.1$ and $A_1/A_2= 1.2 \pm 0.2$, showing a small branch mixing. Over $0\leq\theta_{\rm mw}\leq2\pi$, $r_a$ varies approximately between 0.7 and 1.4. Day-to-day drifts of beam alignment and sheet potential may change the ring profile $|\varphi(\vec{r})|^2$, while branch mixing remains small with variations, $r_a$ varying up to $30\%$ for a given $\theta_{\rm mw}$. For a fixed $\theta_{\rm mw}$, the torus shape varies slightly from day to day, while $\mu_{1,2}$ remain stable within 1.5 standard deviations owing to phase locking. The two-peak modulation persists despite experimental noise in phase, coupling strength, as well as ring trap inhomogeneity, which introduces peak asymmetry without altering the twofold structure. Nevertheless, the fixed total flux of $4\pi$ through the toroidal surface ensures two density maxima, highlighting the robust topological origin of the modulation.
%

The emergence of a two-peak density modulation signifies the formation of a synthetic Hall torus. We further control the peak positions by tuning the microwave phase $\theta_{\rm mw}$, corresponding to the axial flux threading the torus. We repeat the same experimental procedure for various $\theta_{\rm mw}$, where $\theta_{\rm mw}$ is time-independent in each single experiment and $\theta_{\rm mw}$ is varied cross different experiments. Figure~3(ab) shows ${\rm OD}_{m_F=1}$ and the corresponding $\mu_{1,2}$ as $\theta_{\rm mw}$ varies from $0$ to $2\pi$. The azimuthal modulation rotates with $\theta_{\rm mw}$: $\mu_{1,2}$ increases linearly with slope $0.51\pm 0.01$. Thus, advancing $\theta_{\rm mw}$ by $2\pi$ shifts each maximum by $\pi$, requiring a $4\pi$ change in $\theta_{\rm mw}$ to return to its initial position. This reduced periodicity, despite the $2\pi$ periodicity of the Hamiltonian [Eq.(\ref{eq:V})], reflects the underlying nonsymmorphic symmetry of the synthetic Hall torus~\cite{Yan2019}, and is closely related to subwavelength lattice structures~\cite{Anderson2020,Zhang2021, Li2022}.

Whereas the density modulation and its dependence on $\theta_{\rm mw}$ as shown in Fig.~3(ab) readily provide signature of a Hall torus, a direct probe of Hall response requires transport measurement. While Laughlin’s charge pump has been realized on an atomic Hall cylinder~\cite{Fabre2022}, its Hall-torus realization remains experimentally unexplored. To this end, we slowly ramp $\theta_{\rm mw}$ in time to realize a topological charge pump on the torus. Applying $\theta_{\rm mw}(t)=2\pi \times f t$ where $f$ denotes the ramping rate, we track the torus peak positions $\mu_{1,2}$. Figure~3(c) shows $\mu_{1,2}$ versus the scaled time $ft=\theta_{\rm mw}(t)/2\pi$ for $f=5$, 10, and 20~Hz. For sufficiently slow ramps ($f\leq 20$~Hz), all three data sets collapse onto a single curve. This collapse confirms that the dynamics are adiabatic and insensitive to the ramp rate within this regime, as expected for a topological charge pump. Accordingly, changing the axial flux by $2\pi$ shifts each torus branch by $\pi$ along $\phi$, independent of the details of the flux-threading process. Linear fits of $\mu_{1,2}$ versus $ft=\theta_{\rm mw}(t)/2\pi$ give both slopes of $0.50\pm 0.01$, as expected.

\begin{figure}
	\centering
	\includegraphics[width=3.5 in]{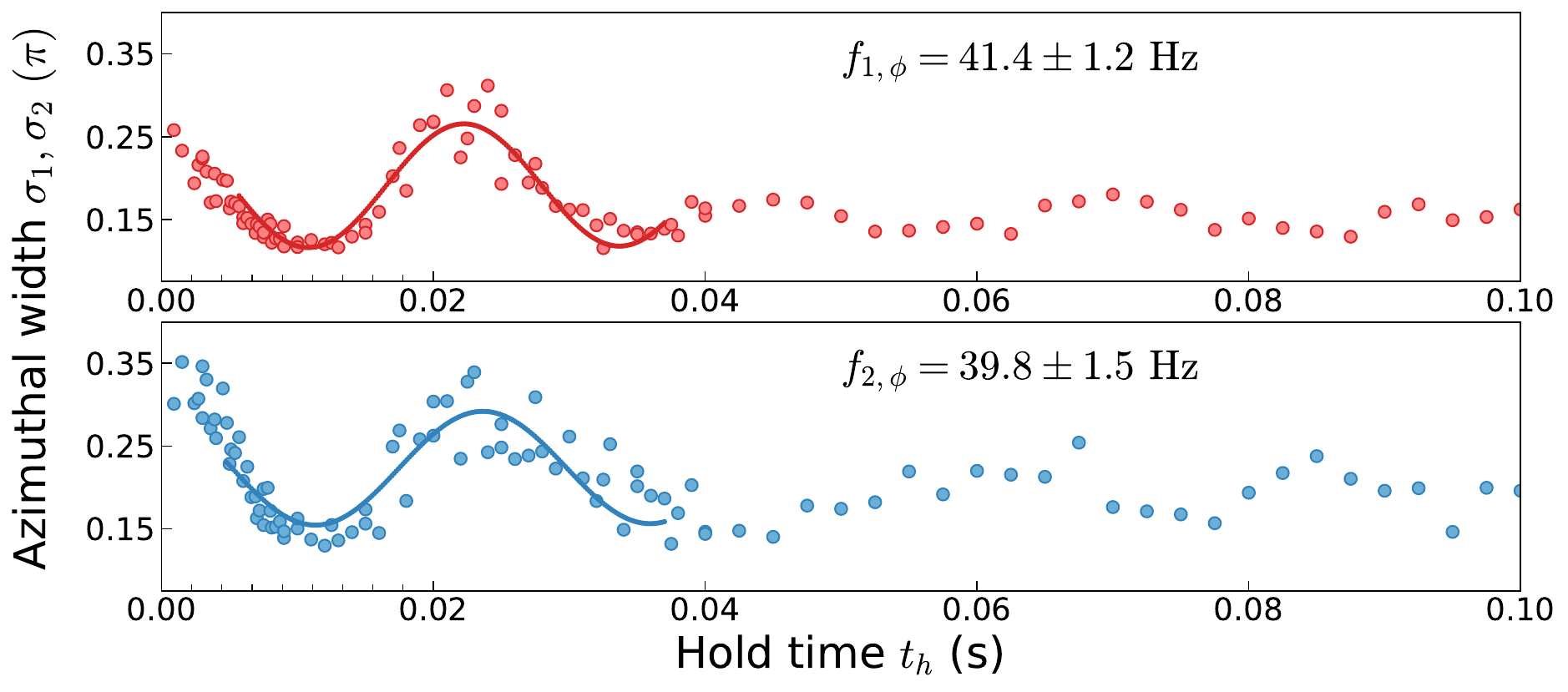}
	\caption{Dynamics after a sudden transition from a cylindrical to a toroidal geometry by abruptly turning on $\Omega_{\rm mw}$. Azimuthal widths $\sigma_1,\sigma_2$ of the torus peaks centered at $\phi=\mu_1,\mu_2$ vs. hold time $t_h$ are displayed in the top and bottom panel, respectively.}
\end{figure}

The pronounced density modulations observed on a Hall torus, absent in a cylindrical geometry (open boundary in the synthetic dimension), motivate us to study their dynamical emergence after a boundary-condition quench. We abruptly turn on the microwave coupling ($t_{\rm on} = 0.01$~ms, $\theta_{\rm mw}=0$), switching from a cylinder, periodic in $\phi$ and open in the synthetic dimension, to a torus with periodic boundary conditions in both directions. We monitor $\tilde{n}_{m_F=1}$ for hold times $t_h$ up to 0.1~s. The initial profile evolves into a two-peak azimuthal modulation with near-zero minima after $t_h\sim 6$~ms, followed by density oscillates due to the mismatch between the initial state and the torus eigenstate. Figure~4 shows the azimuthal widths versus $t_h$. At early times ($t_h\lesssim0.04$~s), the oscillations are nearly sinusoidal, with frequencies $41.4\pm1.2$ and $39.8\pm1.5$~Hz. The quench dynamics can transfer some atoms to excited bands; although their populations cannot be measured directly, the oscillation frequency is reasonably close to the ground-to-first-excited-band gap of $\sim 35$~Hz. At longer times, the dynamics deviate from sinusoidal behavior and decay in amplitude, reflecting complex relaxation dynamics (see Supplemental Material~\cite{supplemental}).

We have demonstrated the first experimental realization of a synthetic Hall torus, observed geometry-induced azimuthal density modulations, implemented a toroidal charge pump, and probed dynamics after a sudden topology change. Our approach provides a platform for exploring quantum physics in curved spaces. More complex geometries, such as spheres, can be realized by replacing the cyclic $|1,1\rangle$--$|1,-1\rangle$ microwave coupling with two additional states, $|2,\pm 1\rangle$, that do not experience the Raman-induced ring trap. Microwave couplings $|1,1\rangle$--$|2,1\rangle$ and $|1,-1\rangle$--$|2,-1\rangle$ then cap the cylinder ends, yielding a closed surface threaded by $4\pi$ flux, topologically equivalent to a sphere enclosing a charge-2 monopole.
While this work focuses on weak interactions, extending it to strongly interacting regime could enable fractional quantum Hall states and topological order on nontrivial manifolds. Dynamically tunable topology and gauge fields further open opportunities for exploring quantum transport and nonequilibrium topological phenomena.

The authors thank I.~B. Spielman, Y. Kawaguchi, C. Chin, C.-W. Chou and K.-Y. Lin for useful discussions. We also thank Y. Yan for valuable discussions and for kindly sharing Gross–Pitaevskii (GP) results on branch mixing and quench dynamics. Y.-J.L. was supported by NSTC 108-2112-M-001-033-MY3 and 111-2112-M-001-048-MY3 and the Academia Sinica Thematic Research Program AS-TP-109-M09. Q.~Z. was supported by the Air Force Office of Scientific Research under Award Number FA9550-20-1-0221.


\widetext
\clearpage
\begin{center}
	\textbf{\large Supplemental Material of ``Realization of a Synthetic Hall Torus with a Topological Charge Pump in a Spinor Bose-Einstein Condensate''}
\end{center}
\setcounter{equation}{0}
\setcounter{figure}{0}
\setcounter{table}{0}
\setcounter{page}{1}
\makeatletter
\renewcommand{\theequation}{S\arabic{equation}}
\renewcommand{\thefigure}{S\arabic{figure}}
\renewcommand{\bibnumfmt}[1]{[S#1]}
\renewcommand{\citenumfont}[1]{S#1}

\maketitle

\section{Formalism of the BEC dressed by Raman and microwave fields}
The Hamiltonian in the bare spin $\ket{m_F=1,0,-1}$ basis under rotating wave approximation in the $(r,\phi,z)$ coordinate is
\begin{align}\label{eq:H}
	\hat{H}=\left[\frac{-\hbar^2}{2m}\frac{\partial}{r\partial
		r}(r \frac{\partial}{\partial
		r})-\frac{\hbar^2}{2m}\frac{\partial^2}{\partial z^2}
	+\frac{\hat{\ell}^2}{2m r^2}\right] \otimes {\hat 1}+\hat{H}_{0},
\end{align}
where $\hat{\ell}=-i\hbar\partial_{\phi}$ is the orbital angular momentum operator, and the atom-light coupling is
\begin{align}\label{eq:Val}
	\hat{H}_{0}/\hbar &= e^{i\Delta \omega_L t}\frac{\Omega(r)}{\sqrt{2}}e^{i\phi} \ket{1}\bra{0} + e^{i\Delta \omega_L t}\frac{\Omega(r)}{\sqrt{2}}e^{i\phi} \ket{0}\bra{-1} \nonumber\\
	&\quad + e^{-i\Delta \omega_{\rm mw} t}\frac{\Omega_{\rm mw}}{2}e^{-i\theta_{\rm mw}} \ket{-1}\bra{1} + \text{H.c.} \sum_{m_F=0,\pm1} \omega_{m_F} \ket{m_F}\bra{m_F},
\end{align}
$\Delta \omega_L, \Delta \omega_{\rm mw}$ are the frequency differences between the two Raman laser beams and between two microwave fields, respectively.
The $\ket{m_F}$ spin state has energy $\hbar\omega_{m_F}$ including the Zeeman energy shifts and the microwave-induced light shifts $\delta \omega_{m_F}$, where $\omega_1=-\omega_Z^{(0)}+\delta \omega_1,\omega_0=-\omega_q^{(0)}+\delta \omega_0,\omega_{-1}=\omega_Z^{(0)}+\delta \omega_{-1}$; $\omega_Z^{(0)}$ and $\omega_q^{(0)}$ are the linear and quadratic Zeeman energy shift, respectively. We further define the effective Zeeman shift $\omega_Z=(\omega_{-1}-\omega_1)/2$ and the effective quadratic Zeeman shift $\omega_q=(\omega_{-1}+\omega_1)/2-\omega_0$. Then, we rewrite $\omega_{m_F}$ by adding an energy offset, and express as
\begin{align}
	\hat{H}_{0}=\hbar\left(
	\begin{array}{ccc}
		-\omega_Z & e^{i\Delta \omega_L t}\frac{\Omega(r)}{\sqrt{2}}e^{i\phi} & e^{i\Delta \omega_{\rm mw} t}\frac{\Omega_{\rm mw}}{2}e^{i\theta_{\rm mw}}\\
		e^{-i\Delta \omega_L t}\frac{\Omega(r)}{\sqrt{2}}e^{-i\phi} & -\omega_q & e^{i\Delta \omega_L t}\frac{\Omega(r)}{\sqrt{2}}e^{i\phi}\\
		e^{-i\Delta \omega_{\rm mw} t}\frac{\Omega_{\rm mw}}{2}e^{-i\theta_{\rm mw}} & e^{-i\Delta \omega_L t}\frac{\Omega(r)}{\sqrt{2}}e^{-i\phi} & \omega_Z
	\end{array}\right).
\end{align}

In our experiment, we set the microwave frequency difference $\Delta \omega_{\rm mw}$ equal to twice of the Raman frequency difference $\Delta \omega_L$,
\begin{align}
	2\Delta \omega_L-\Delta \omega_{\rm mw}=0.
\end{align}
We then express $\hat{H}_{0}$ in the frame rotating at $\Delta\omega_L$, which is the general form of $V$ in our main text,
\begin{align}
	V=\hbar\left(
	\begin{array}{ccc}
		\delta &\frac{\Omega(r)}{\sqrt{2}}e^{i\phi} & \frac{\Omega_{\rm mw}}{2}e^{i\theta_{\rm mw}}\\
		\frac{\Omega(r)}{\sqrt{2}}e^{-i\phi} & -\omega_q & \frac{\Omega(r)}{\sqrt{2}}e^{i\phi}\\
		\frac{\Omega_{\rm mw}}{2}e^{-i\theta_{\rm mw}} & \frac{\Omega(r)}{\sqrt{2}}e^{-i\phi} & -\delta
	\end{array}\right),
\end{align}
where the Raman detuning $\delta=\Delta \omega_L-\omega_Z$ and $V$ is time-independent.

\subsection{Raman dressed state light shift}
We consider the case when the microwave coupling $\Omega_{\rm mw}$ is zero. These Raman beams introduce the coupling to the atoms in the frame rotating at frequency $\Delta \omega_L$, 
\begin{align}
	H_{\Omega}/\hbar= \frac{\Omega(r)}{\sqrt{2}}e^{i\phi} \ket{1}\bra{0} + \frac{\Omega(r)}{\sqrt{2}}e^{i\phi} \ket{0}\bra{-1}\pm \delta|\pm1\rangle\langle \pm1|+\text{H.c.}=\vec{\Omega}_\textrm{eff}(\vec{r})\cdot \vec{F},
\end{align}
where $\vec{F}$ is the vector of spin operators and $\vec{\Omega}_\textrm{eff}=\Omega(r)\cos\phi {\bf e}_x-\Omega(r)\sin\phi {\bf e}_y+\delta {\bf e}_z$ with  $\delta$ being the Raman detuning and $\Omega(r)=\Omega_M\sqrt{e}(r/r_M)e^{-r^2/2r^2_M}$ the Raman coupling strength. Here we ignore the quadratic Zeeman shift for $\Omega_{\rm mw}=0$, $\omega_q^{(0)}/2\pi=50$~Hz.
Because $H_\Omega$ dominates the spin-dependent part of the Hamiltonian, it is convenient to introduce the dressed spin basis $|\xi_{n=0,\pm1}(\vec{r})\rangle$ defined by $H_\Omega|\xi_n(\vec{r})\rangle=n|\vec{\Omega}_\textrm{eff}(\vec{r})||\xi_n(\vec{r})\rangle$, where $|\xi_n(\vec{r})\rangle$ is a normalized spinor part of the wave function. The overall spinor order parameter for atoms being loaded the the lowest energy dressed state is $\varphi(\vec{r})|\xi_{-1}(\vec{r})\rangle$ with a scalar wave function $\varphi(\vec{r})$, and the eigenenergy is  $\Omega_{\textrm{eff},-1}=-|\vec{\Omega}_\textrm{eff}(\vec{r})|$,
\begin{align}
	\Omega_{\textrm{eff},-1}= -\sqrt{\Omega(r)^2+\delta^2}.	
\end{align}
The dressed state light shift for $\ket{\xi_{-1}}$ is $-\sqrt{\Omega(r)^2+\delta^2}$, which is $-\Omega(r)^2/2|\delta|$ up to a constant for large $|\delta|$ and is $-\Omega(r)$ for $\delta=0$. Thus, in our Raman dress state before the microwave is turned on, the dressed state light shift $-\Omega(r)$ provides a ring-shaped potential.

\section{Experimental parameters and procedures}
At the beginning of the experiment, we perform evaporative cooling on $^{87}$Rb atoms in the state $\ket{F, m_F} = \ket{1, -1}$, confined in a crossed optical dipole trap~\cite{SChen2018a} combined with a 767 nm blue-detuned sheet potential. The 767 nm light is provided by a Toptica TA pro system consisting of a narrow-linewidth master laser and a tapered amplifier. The sheet beam propagates along ${\mathbf e}_y$ and has a Hermite-Gaussian $\mathrm{TEM}_{01}$ profile along ${\mathbf e}_z$, providing tight vertical confinement. In addition, the sheet beam introduces weak anti-trapping potentials along ${\mathbf e}_x$ and ${\mathbf e}_y$.

We then adiabatically load a cloud of cold thermal atoms into the lowest energy spinor branch Raman-dressed state $\ket{\xi_{-1}}$~\cite{SChen2018a} with detuning $\delta=0$. After loading, the atoms have temperature $T\gtrsim T_c$, where $T_c$ is the BEC critical temperature. We continue evaporative cooling by reducing the dipole trap power to near zero over $0.3$~s, yielding a condensate of $N = (8 \pm 1) \times 10^4$ atoms in the Raman-dressed state. The dressed state light shift at $\delta=0$ provides a ring-shaped trapping potential in the $xy$ plane. In addition, the crossed dipole trap provides transverse confinement with $\omega_r/2\pi=25$~Hz, and the overall trap frequency along ${\mathbf e}_z$ is about 370~Hz.

The two Raman beams co-propagate along the ${\mathbf e}_z$ direction: one is Gaussian (G), and the other is Laguerre-Gaussian (LG). Both beams operate at $\lambda = 790$ nm, where the scalar light shifts from the D1 and D2 lines cancel. The LG and G beams have optical frequencies $\omega_L$ and $\omega_L + \Delta \omega_L$, and are linearly polarized along ${\mathbf e}_x$ and ${\mathbf e}_y$, respectively. The LG beam, generated using a vortex phase plate, carries a phase winding number $1$ and has radial index zero. The foci of both beams are aligned near the BEC position along ${\mathbf e}_z$, with the G beam having a waist of $\approx 200~\mu$m. The LG beam center is nearly coincident with the dipole trap center at $(r = 0, z = 0)$, with an estimated deviation of $\lesssim 2\micron$. 

The two-photon microwave coupling is produced by $\ket{1,-1}\rightarrow\ket{2,0}$ and $\ket{1,1}\rightarrow\ket{2,0}$ where $\ket{2,0}$ is the intermediate state and one-photon detuning is $\Delta/2\pi=100$~kHz; the respective frequencies are $\omega_{{\rm mw},-}$ and $\omega_{{\rm mw},+}=\omega_{{\rm mw},-}+\Delta\omega_{\rm mw}>\omega_{{\rm mw},-}$ with one-photon Rabi frequencies $\Omega_-/2\pi\approx\Omega_+/2\pi\approx 25$~kHz. These parameters give the two-photon microwave coupling strength $\Omega_{\rm mw}/2\pi\approx3.15$~kHz for the synthetic torus. The linear Zeeman shift without microwave fields for $F=1,2$ manifolds is $\omega_Z^{(0)}/2\pi\sim 550$~kHz, which is more than 5 times the one-photon detuning $\Delta/2\pi$. Therefore, we consider four dominate coupling routes, $\ket{1,-1}\rightarrow\ket{2,0}$, $\ket{1,1}\rightarrow\ket{2,0}$, $\ket{1,0}\rightarrow\ket{2,-1}$ and $\ket{1,0}\rightarrow\ket{2,1}$, each of which has one-photon detuning $\Delta/2\pi=100$~kHz. We consider the effects of off-resonant one-photon coupling as below: the $\ket{1,0}$ state is coupled to both $\ket{2,-1}$ and $\ket{2,1}$ with admixtures amplitude of $\Omega_-/2\Delta,\Omega_+/2\Delta$, leading to population in $\ket{2,0}$ of $2(\Omega_{\mp}/2\Delta)^2\sim~3\%$. Besides, we consider coupling within $\ket{1,\pm1}$ and $\ket{2,0}$, where the amplitude of $\ket{2,0}$ is $c_{2,0}\sim -(\Omega_+ c_{1,1}+\Omega_- c_{1,-1})/2\Delta$ and the population in $\ket{2,0}$ is $c_{2,0}^2\sim~3\%$ for $c_{1,1}\sim c_{1,-1}\sim1/\sqrt{2}$. Here, the off-resonant one-photon coupling induces unwanted population disturbance in $m_F=0$ at the few-percent level. This is not an extremely deep large-detuning regime owing to the limit on the experimental microwave power.

We derive the effective Zeeman and quadratic Zeeman shifts by measuring microwave-induced differential light shifts between $\ket{m_F}$ and $\ket{m_F-1}$, which are $\delta \omega_1-\delta \omega_0$ and $\delta \omega_{0}-\delta \omega_{-1}$ and are at the order of $2\pi\times$~kHz, with microwave two-photon resonance spectroscopy. In the time sequence of turning on the microwave coupling, we ramp $\Omega_{\rm mw},\omega_q$ with a weak quadratic function $\Omega_{\rm mw}/(2\pi\times 3.15~{\rm kHz})=\omega_q/(2\pi\times 1.6~{\rm kHz})=-0.14f^2+1.14f$ for $0\leq f=t/t_{\rm on}\leq 1$; the simultaneous ramp of bias field compensates the light shifts of microwave coupling such that the detuning $|\delta/2\pi|$ is within $\sim 150$~Hz.

Here we discuss spin miscibility of $\ket{F=1,m_F=\pm1,0}$. The spin-dependent interaction of $^{87}$Rb atoms is rather weak and is insufficient to create measurable spatial separation between different spin states or modified density distribution in our system under Raman and microwave coupling $\Omega_M,\Omega_{\rm mw}$, respectively. The parameters for spin-independent and spin-dependent part of the interactions are $\tilde{c}_0=4\pi\hbar^2(a_0+2a_2)/3m$, and $\tilde{c}_2=4\pi\hbar^2(a_2-a_0)/3m<0$, respectively, where $a_f$ is the s-wave scattering length in the total spin $f$ channel. The ratio $\tilde{c}_2/\tilde{c}_0=-0.0046$ is rather small, signifying the weakness of the spin-dependent interaction. Furthermore, using realistic density $n$ in experiments, the spin-dependent interaction energy $|n\tilde{c}_2|$ is about 3~Hz, about three orders of magnitude smaller than the strengths of the Raman and microwave couplings $\Omega_M/2\pi,\Omega_{\rm mw}/2\pi\sim 3$~kHz. All the three spin components are thus well mixed. This is the reason that in both experiments [Fig.~2(ac)] and simulations [Fig.~S1(a)], we find all three spin components have the same density-modulated spatial profiles, which directly shows the three spin components are spatially miscible.

We use a microwave field to selectively pump the atoms from $\ket{F=1,m_F}$ to $\ket{F=2}$ and perform resonant absorption imaging of $F=2 \rightarrow F^{\prime}=3$ transition. Each $|m_F\rangle$ image is obtained in an individual experimental realization. All the images shown in the paper are single-shot.

In our experiments, several factors can contribute to cylindrical asymmetry. For a BEC centered at $(r=0,z=0)$, both Raman beams can be slightly misaligned in the transverse $xy$ plane or have focal positions shifted along ${\mathbf e}_z$, introducing additional phase shifts due to the curvature of their wavefronts. The blue-detuned 767 nm sheet beam also produces a weak anti-trapping potential in the $xy$ plane, with different curvatures along ${\mathbf e}_x$ and ${\mathbf e}_y$. Additionally, we observe that the ring trap profile varies with the current of the 767 nm master laser, while the laser remains single frequency mode with fixed optical power. This suggests the presence of additional, unmodeled potentials from the sheet beam can break cylindrical symmetry.

\section{Experimental data analysis}
Our experimental and simulation data of the atomic $\mathrm{OD}$ are in cartesian coordinate in discrete pixels. We first use an interpolation to obtain $\mathrm{OD}(x,y)$ for continuous variables $(x,y)$, and then analyze the data in cylindrical coordinate $(r,\phi)$. 

The true 1D atomic density of spin $m_F$ state is given by
\begin{align}
	n_{m_F}^{\rm 1D}(\phi-\theta_{\rm mw}/2)=\int dr r n_{m_F}^{\rm 2D}(r,\phi-\theta_{\rm mw}/2)=\bar{\sigma}^{-1}\int dr r {\mathrm{OD}}_{m_F}(r,\phi-\theta_{\rm mw}/2),
\end{align} 
where the 2D density $n_{m_F}^{\rm 2D}=\bar{\sigma}^{-1}{\mathrm{OD}}_{m_F}$ is given by the atomic OD and the imaging cross section $\bar{\sigma}$. Since the 1D density $n_{m_F}^{\rm 1D}$ is proportional to the integrated OD along $r$, we then define the effective 1D density as
\begin{align}
	\tilde{n}_{m_F}(\phi)=\frac{1}{A_0}\Big[\int_{\phi-\Delta \phi/2}^{\phi+\Delta \phi/2} d\phi^{\prime}\Big]^{-1}\int_{\phi-\Delta \phi/2}^{\phi+\Delta \phi/2} d\phi^{\prime}\int_{r}dr r \mathrm{OD}(r,\phi^{\prime}),
\end{align}
where the OD is integrated within the azimuthal angle $\phi-\Delta \phi/2<\phi^{\prime}<\phi+\Delta \phi/2$ and is averaged within the binning angle $\Delta\phi$, and $A_0$ is the area of a single pixel. The pixel sizes are $1.2~\mu\mathrm{m}$ for experimental data and $0.5596~\mu\mathrm{m}$ for simulations, and the typical binning angle is $\Delta\phi=0.056\pi$.

\section{Numerical simulations}

Gross–Pitaevskii simulations included in this section were produced by Prof. Y. Yan at The Chinese University of Hong Kong, who has generously shared with us his results related to a previous work~\cite{SLi2022}.

\subsection{Torus branches and their mixing}
In the ideal case of a cylindrically symmetric system, where the Raman beams and the resulting ring trap are perfectly symmetric and centered on the BEC, 
imaginary-time propagation was used to compute the 3D Gross–Pitaevskii (GP) ground states of the torus for branches 1 and 2 at $\ell = 0$, denoted as $|\Phi_1\rangle$ and $|\Phi_2\rangle$, respectively~\cite{SLi2022}. Since the energy $E_0\approx h\times 0.3$~Hz is much smaller than the Raman and microwave coupling strength, both branches have nearly the same spinor number fraction of $(0.235, 0.529,0.235)$ in $m_F=(1,0,-1)$ state. Both branches have the same density profile, as shown by the spinor optical density profiles of $|\Phi_1\rangle$ in Fig.~S1(a).

To gain further insight into the asymmetry between the two torus peaks observed in the experiment, 
in the GP simulations one needs to include possible misalignments of the Raman beams relative to the BEC center—an imperfection we believe to be a primary contributor. While it is difficult to account for all experimental sources of asymmetry (see Sec. ``Experimental setup and procedures'' for details on the setup), this modeled misalignment allows us to explore its impact. However, the simulated density profiles do not fully reproduce the asymmetries observed in the experiment. This discrepancy shows the challenge of faithfully capturing the experimental torus shapes in simulations, given the presence of uncontrolled or unmodeled imperfections.

\begin{figure} 
	\centering\includegraphics[width=6.0 in]{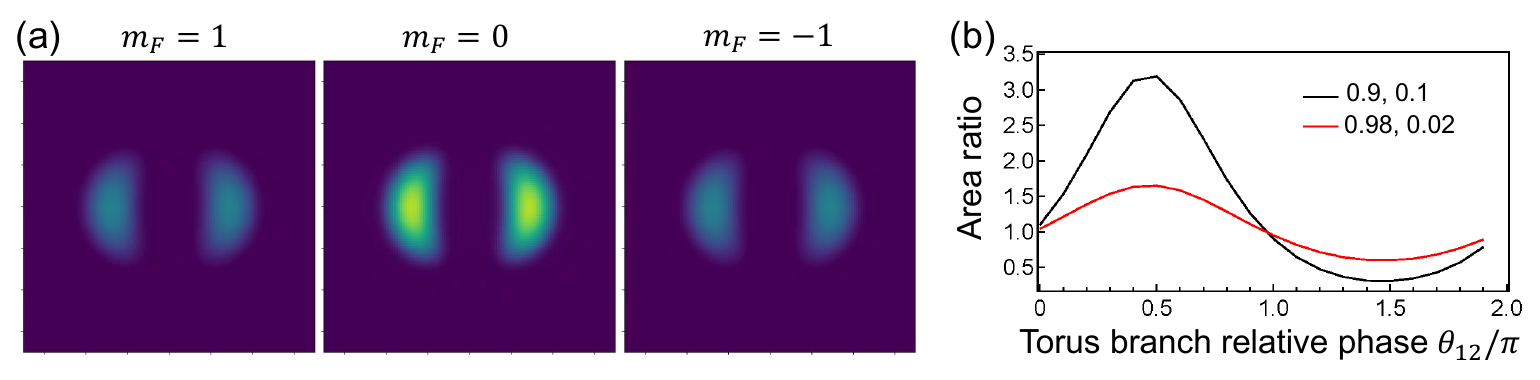} 
	\caption{(a) Optical density profiles of the torus in branch 1 for the ideal system with a perfect cylindrical symmetry. Branch 2 has identical density profiles as branch 1. (b) The area ratio $r_a$ as a function of the relative phase $\theta_{12}$ in the mixing of the two ground state branches of the torus. Branch mixtures of $(0.9,0.1)$ and $(0.98,0.02)$ are displayed.
	}
\end{figure}

Cylindrical asymmetry is one of the two factors that induces mixing between the two torus branches $|\Phi_1\rangle$ and $|\Phi_2\rangle$. The amplitude of this mixing depends on the azimuthal structure of the external wave function $\varphi(\vec{r})$ in the ring trap (before we turn on the microwave coupling), as well as on the microwave phase $\theta_{\rm mw}$. Rather than focusing on the detailed ring trap potential, we examine the torus peak area ratio $r_a$ from the azimuthal density profiles $\tilde{n}_{m_F=1}$ as superpositions of $|\Phi_1\rangle$ and $|\Phi_2\rangle$ with varying weights and relative phases $\theta_{12}$, where $r_a\neq1$ indicates asymmetry between the two peaks. As shown in Fig.~S1(b) with branch population fractions of $(0.9,0.1)$ and $(0.98,0.02)$, which correspond to mixing ratios of $10\%,2\%$, respectively, even a small admixture of the two branches can produce a noticeable asymmetry between the two density peaks—qualitatively consistent with experimental observations. The area ratio $r_a$ oscillates with $\theta_{12}$ and the maximum of $r_a>1$ increases with the mixing ratio. The 0.02 mixing ratio gives a similar range of $r_a$ to that in the experiment.

\subsection{Quench dynamics}
The quench dynamics were simulated by solving the time-dependent Gross–Pitaevskii equation (TDGPE), assuming a perfectly cylindrically symmetric system for simplicity. The initial state is the GP ground state of Raman dressed state with $\tilde{\ell}=0$, in the absence of the microwave field. At time $t_h=0$, the microwave coupling $\Omega_{\rm mw}$ is suddenly turned on, loading the atoms into the torus branch 1. We track the evolution of the azimuthal root-mean-square widths $\sigma_{1,2}$ of the torus peaks versus hold time $t_h$ as torus features emerge, where we find $\sigma_1\approx \sigma_2$ as expected. 

\begin{figure} 
	\centering\includegraphics[width=5.5 in]{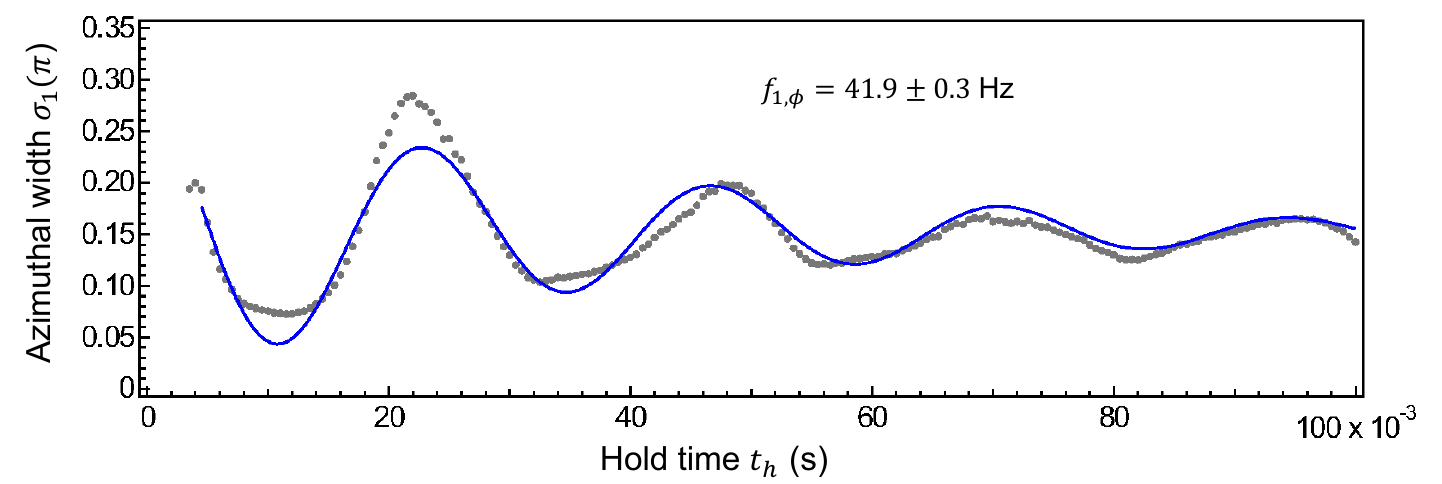}
	\caption{Simulations of azimuthal width $\sigma_1$ vs. hold time $t_h$ after quench for a cylindrically symmetric system. Dark gray symbols (blue curve) denote the numerical data (fit). Unlike the experimental data in the main text, numerical results can well fit to a damped sinusoidal oscillation even at long times.
	}
\end{figure}

Figure.~S2 shows $\sigma_{1}$ as a function of $t_h$, along with a fit to a damped sinusoidal oscillation. While the fit shows some deviation from the numerical data, it yields a characteristic frequency of approximately 42~Hz. At short time $t_h\lesssim 0.04$~s, the numerical results are close to the experimental data presented in Fig.~4 of the main text, showing similar oscillation frequencies. However, at longer $t_h$, the irregular dynamics observed in the experiment are no longer captured by the TDGPE simulation and cannot be well fit by a damped sinusoidal oscillation. This discrepancy suggests that a complete understanding of the quench dynamics requires more sophisticated studies of branch mixing, cylindrical asymmetries, and beyond mean-field effects in the system.

\end{document}